\documentclass[11pt]{article}
\usepackage{graphicx}
\usepackage{amsfonts}
\usepackage{theorem}
\newtheorem{Lem}{Lemma}[section]
\newtheorem{Def}[Lem]{Definition}
\newtheorem{The}[Lem]{Theorem}
\newtheorem{Prop}[Lem]{Proposition}
\newtheorem{Cor}[Lem]{Corollary}
\newtheorem{Ex}[Lem]{Example}

\newtheorem{Rem}[Lem]{Remark}

\newcommand{\qed}{\hbox{\rule{6pt}{6pt}}}

\begin{document}
\title{On the maximum entropy principle and the minimization of the Fisher information in Tsallis statistics} 
\author{Shigeru Furuichi$^1$\footnote{E-mail:furuichi@chs.nihon-u.ac.jp}\\
$^1${\small Department of Computer Science and System Analysis, 
College of Humanities and Sciences,}\\{\small Nihon University,
3-25-40, Sakurajyousui, Setagaya-ku, Tokyo, 156-8550, Japan}}
\date{}
\maketitle
{\bf Abstract.} 
We give a new proof of the theorems on the maximum entropy principle in Tsallis statistics.
That is, we show that the $q$-canonical distribution attains the maximum value of the 
Tsallis entropy, subject to the constraint on the $q$-expectation value
and the $q$-Gaussian distribution attains the maximum value of the Tsallis entropy, subject to the constraint on the $q$-variance,
as applications of the nonnegativity of the Tsallis relative entropy, without using the Lagrange multipliers method.
In addition, we define a $q$-Fisher information and then prove a $q$-Cram\'er-Rao inequality that 
the $q$-Gaussian distribution with special $q$-variances attains the minimum value of
the $q$-Fisher information.
\vspace{3mm}


\vspace{3mm}


\vspace{3mm}

\section{Introduction}
The problems on maximum entropy principle in Tsallis statistics \cite{Tsa,AO} has been studied in classical system and quantum system \cite{MNPP,TMP,AMPP,Abe}. 
Such problems have been solved by the use of the Lagrange multipliers formalism. However we give a new proof for such problems, 
that is, we prove them by applying the Tsallis relative entropy without Lagrange multipliers formalism.  
Moreover, we derive a one-parameter extended
Cram\'er-Rao inequality involving a one-parameter extended Fisher information.

We denote the $q$-logarithmic function $\ln_q$ by 
$$\ln_{q}x \equiv \frac{x^{1-q} -1}{1-q} \,\,\,\,\,\,\,\,\,\, (q\in\mathbb{R}, q \neq 1, x > 0)$$
and the $q$-exponential function $\exp_q$ by
\[
\exp _q \left( x \right) \equiv \left\{ \begin{array}{l}
 \left( {1 + \left( {1 - q} \right)x} \right)^{\frac{1}{{1 - q}}} ,\,\,\,\,\,\,if\,\,\,1 + \left( {1 - q} \right)x > 0, \\ 
 \,\,\,\,\,\,\,\,\,\,\,\,\,\,\,\,\,0\,\,\,\,\,\,\,\,\,\,\,\,\,\,\,\,\,\,\,\,\,\,\,\,\,\,\,\,\,\,\,\,\,\,\,\,\,otherwise \\ 
 \end{array} \right.\,\,\,\,\,\,\,\,\,\,\,\,\,\,\left( {q \in \mathbb{R},q \ne 1,x \in \mathbb{R}} \right).
\]
The functions $\exp_{q}(x) $ and $\ln_{q}x$ converge to $\exp(x)$ and $\log x$ as $q \to 1$, respectively.
Note that we have the following relations:
\begin{equation}
 \exp_{q}\left\{x+y+(1-q) xy\right\} = \exp_{q}(x) \exp_{q}(y), \,\,\,\,\,
 \ln_{q} xy = \ln_{q}x + \ln_{q} y + (1-q) \ln_{q} x \ln_{q} y. \label{nonex01}
\end{equation}

In the following of this section, we define the Tsallis entropy and the Tsallis relative entropy 
for the probability density functions.
The set of all probability density function on $\mathbb{R}$ is represented by 
$$ D\equiv \left\{ f :\mathbb{R} \to \mathbb{R} : f(x) \geq 0, \int_{-\infty}^{\infty} f(x) dx =1\right\}.$$
Then the Tsallis entropy \cite{Tsa} is defined by 
\begin{equation}
H_{q}(\phi(x)) \equiv - \int_{-\infty}^{\infty} \phi(x)^q \ln_{q} \phi(x) dx
\end{equation}
for any nonnegative real number $q\neq 1$ and a probability density function $\phi(x) \in D$. 
In addition, the Tsallis relative entropy is defined by
\begin{equation}
D_q(\phi(x)\vert \psi(x)) \equiv \int_{-\infty}^{\infty} \phi(x)^q (\ln_q\phi(x) -\ln_q \psi(x))dx
\end{equation}
for any nonnegative real number $q \neq 1$ and two probability density functions $\phi(x)\in D$ and $\psi(x)\in D$. 
Taking the limit $q \to 1$, the Tsallis entropy and the Tsallis relative entropy converge to 
the Shannon entropy $H_1(\phi(x)) \equiv -\int_{-\infty}^{\infty} \phi(x) \log \phi(x)dx$ and the Kullback-Leibler divergence
$D_1(\phi(x)\vert \psi(x)) \equiv \int_{-\infty}^{\infty} \phi(x) (\log \phi(x) -\log \psi(x))dx$, respectively.
See \cite{FYK} for fundamental properties on the Tsallis relative entropy.

We define two sets involving the constraints on the normalized $q$-expectation value and $q$-variance:
$$
C_q^{(c)} \equiv \left\{ f \in D:   \frac{1}{c_q}\int_{-\infty}^{\infty}x f(x)^qdx = \mu_q    \right\}
$$
and
$$
C_q^{(g)} \equiv \left\{ f \in C_q^{(c)} :    \frac{1}{c_q}\int_{-\infty}^{\infty}(x - \mu_q)^2f(x)^qdx = \sigma_q^2     \right\},
$$
where $c_q \equiv \int_{-\infty}^{\infty} f(x)^q dx$ is a normalization factor.

Then the $q$-cannonical distribution $\phi_q^{(c)}(x) \in D$ and the 
$q$-Gaussian distribution $\phi_q^{(g)}(x)\in D$ were formulated in \cite{MNPP,TMP,AMPP,Abe,Suy,ST} by
$$
\phi_q^{(c)}(x) \equiv \frac{1}{Z_q^{(c)}} \exp_q\left\{ -\beta_q^{(c)} (x-\mu_q) \right\}, \,\,\,
Z_q^{(c)} \equiv \int_{-\infty}^{\infty} \exp_q\left\{ -\beta_q^{(c)} (x-\mu_q) \right\}
$$
and
$$
\phi_q^{(g)}(x) \equiv \frac{1}{Z_q^{(g)}} \exp_q\left\{ -\frac{\beta_q^{(g)}(x-\mu_q)^2}{\sigma_q^2} \right\}, \,\,\, 
Z_q^{(g)} \equiv \int_{-\infty}^{\infty} \exp_q\left\{ -\frac{\beta_q^{(g)}(x-\mu_q)^2}{\sigma_q^2} \right\},
$$
respectively, where $\beta_q^{(c)}$ and $\beta_q^{(g)}$ are constant numbers depending on the parameter $q$, and we often use $\beta_q^{(g)}=\frac{1}{3-q}$.

\section{Tsallis maximum entropy principle}\label{sec2}
In this section, we revisit the maximum entropy principle in nonextensive statistical physics. 
The maximum entropy principles in Tsallis statistics have been studied and modified in many literatures \cite{MNPP,TMP,AMPP,Abe,Suy}. 
Here we prove two theorems that maximize the Tsallis entropy under 
two different constraints by the use of the nonnegativity of the Tsallis relative entropy instead of the use of the Lagrange multipliers method.
\begin{Lem}
For $q \neq 1$, we have
$$D_{q} (\phi(x) \vert \psi(x)) \geq 0,$$
with equality if and only if $\phi(x) = \psi(x)$ for all $x$.
\end{Lem}
{\it Proof}:
Since we have $\ln_q x \leq x-1$ with equality if and only if $x=1$ for any $q\in\mathbb{R},q\neq 1$, we have
$$
D_q(\phi(x)\vert \psi(x)) =-\int_{-\infty}^{\infty}
\phi(x)\ln_q\frac{\psi(x)}{\phi(x)}dx \geq -\int_{-\infty}^{\infty}
\phi(x)\left(\frac{\psi(x)}{\phi(x)}-1\right)dx =0,
$$
with equality if and only if $\phi(x) = \psi(x)$ for all $x$.
\hfill \qed

\begin{The} \label{the1}
If $\phi \in C_q^{(c)}$, then $$H_{q}(\phi(x)) \leq -c_q\ln_q   \frac{1}{Z_q^{(c)}},$$ with equality
 if and only if $$\phi(x) = \frac{1}{Z_q^{(c)}}  \exp_q\left\{ -\beta_q^{(c)} (x - \mu_q)  \right\},$$ 
where $\beta_q^{(c)}$ is a constant number depending on the parameter $q$, 
$Z_q^{(c)} \equiv \int_{-\infty}^{\infty}  \exp_q\left\{ -\beta_q^{(c)} (x - \mu_q)  \right\}dx$ and $c_q \equiv \int_{-\infty}^{\infty} \phi(x)^qdx$. 
\end{The}
{\it Proof}:
Putting 
\[
\psi \left( x \right) = \frac{1}{{Z_q^{(c)} }}\exp _q \left(  - \beta _q^{(c)}  \left( x - \mu_q  \right) \right),
Z_q^{(c)}  \equiv \int_{ - \infty }^\infty  {\exp _q \left( -\beta_q^{(c)} \left( x - \mu_q  \right) \right)dx} 
\]
and taking an account for 
$\ln _q \frac{y}{x} = \ln _q y + y^{1 - q} \ln _q \frac{1}{x}$
 and 
$\ln _q \frac{1}{x} =  - x^{q - 1} \ln _q x$,
we have
\begin{eqnarray*}
&& \int_{ - \infty }^\infty  {\phi \left( x \right)^q \ln _q \psi \left( x \right)dx}  = 
 \int_{ - \infty }^\infty  {\phi \left( x \right)^q \ln _q \left\{ {\frac{1}{{Z_q^{(c)} }}\exp _q \left(  - \beta _q^{(c)}  \left( x - \mu_q  \right)  \right)} \right\}dx}  \\ 
 &&= \int_{ - \infty }^\infty  \phi \left( x \right)^q \left\{ -\beta_q^{(c)} \left( x - \mu_q  \right)  
+ \exp _q \left(  - \beta _q^{(c)}  \left( x - \mu_q  \right)  \right)^{1 - q} \ln _q \frac{1}{Z_q^{(c)} } \right\}dx  \\ 
 &&=  - \beta _q^{(c)}  \int_{ - \infty }^\infty  {\left( {x - \mu_q } \right) \phi \left( x \right)^q dx}  + 
\ln _q \frac{1}{{Z_q^{(c)} }}\int_{ - \infty }^\infty  {\phi \left( x \right)^q \left\{ 1 - \beta _q^{(c)} \left( {1 - q} \right)\left( {x - \mu_q } \right) \right\}dx}  \\ 
 &&=   c_q \ln _q \frac{1}{{Z_q^{(c)} }}.
 \end{eqnarray*}
Thus we have 
\[
H_q \left( {\phi \left( x \right)} \right) \equiv  - \int_{ - \infty }^\infty  {\phi \left( x \right)^q \ln _q \phi \left( x \right)dx} 
 \le  - \int_{ - \infty }^\infty  {\phi \left( x \right)^q \ln _q \psi \left( x \right)dx} \, 
=  - c_q \ln _q \frac{1}{{Z_q^{(c)} }}
\]
by the nonnegativity of the Tsallis relative entropy.
From the equality condition of the Tsallis relative entropy,
we see that the maximum attains if and only if 
\[
\phi \left( x \right) = \psi \left( x \right) = \frac{1}{{Z_q^{(c)} }}\exp _q \left(  -\beta_q^{(c)} \left( x - \mu_q  \right)\right).
\]
\hfill \qed

\begin{Rem}
The generalized free energy takes minimum:
$$F_q  \equiv \mu_q - \frac{1}{\beta_q^{(c)}}H_q(\phi(x)) \geq \mu_q + \frac{c_q}{\beta_q^{(c)}}\ln_q\frac{1}{Z_q^{(c)}} $$
if and only if $\phi \left( x \right) = \frac{1}{{Z_q^{(c)} }}\exp _q \left(  -\beta_q^{(c)} \left( x - \mu_q  \right)\right)$
due to Theorem \ref{the1}. 
\end{Rem}

\begin{Cor}\label{cor1}
If $\phi \in C_1^{(c)}$, then $H_{1}(\phi(x)) \leq \log Z_1^{(c)}$
with equality  if and only if  $\phi(x) = \frac{1}{Z_1^{(c)}}    \exp\left\{ -\beta_1^{(c)} (x - \mu) \right\}$.
\end{Cor}
{\it Proof}:
Take the limit $q \to 1$ in Theorem \ref{the1}.
\hfill \qed

By the condition on the existence of $q$-variance $\sigma_q$ 
(i.e., the convergence condition of the integral $\int x^2\exp_q(-x^2)dx$), 
we consider $q \in \mathbb{R}$ such that $0\leq q<3$ and $q\neq 1$.

\begin{The}\label{the2}
For $q\in\mathbb{R}$ such that $0\leq q<3$ and $q\neq 1$ if $\phi \in C_q^{(g)}$, 
then 
$$H_{q}(\phi(x)) \leq  -c_q \ln_q \frac{1}{Z_q^{(g)}} + c_q  \beta_q^{(g)} Z_q^{{(g)}^{q-1}},$$ 
with equality if and only if  
$$\phi(x) = \frac{1}{Z_q^{(g)}}     \exp_q\left( - \frac{\beta_q  ^{(g)} {(x - \mu_q)^2}}{\sigma_q^2}  \right),$$ 
where $Z_q^{(g)} \equiv \int_{-\infty}^{\infty}  \exp_q\left\{ -\beta_q  ^{(g)} (x - \mu_q)^2/{\sigma_q^2}  \right\}dx$
 with $\beta_q^{(g)} = 1/(3-q)$.
\end{The}
{\it Proof}:
Putting 
\[
\psi \left( x \right) = \frac{1}{{Z_q^{(g)} }}\exp _q \left( { - \frac{{\beta _q  ^{(g)} \left( {x - \mu_q } \right)^2 }}{{\sigma_q ^2 }}} \right),
Z_q^{(g)}  \equiv \int_{ - \infty }^\infty  {\exp _q \left( { - \frac{{\beta _q ^{(g)}  \left( {x - \mu_q } \right)^2 }}{{\sigma_q ^2 }}} \right)dx} 
\]
and taking account for 
$\ln _q \frac{y}{x} = \ln _q y + y^{1 - q} \ln _q \frac{1}{x}$
and
$\ln _q \frac{1}{x} =  - x^{q - 1} \ln _q x$,
we have
\begin{eqnarray*}
&& \int_{ - \infty }^\infty  {\phi \left( x \right)^q \ln _q \psi \left( x \right)dx}  = 
 \int_{ - \infty }^\infty  {\phi \left( x \right)^q \ln _q \left\{ {\frac{1}{{Z_q^{(g)} }}\exp _q \left( { - \frac{{\beta _q ^{(g)}  \left( {x - \mu_q } \right)^2 }}{{\sigma_q ^2 }}} \right)} \right\}dx}  \\ 
 &&= \int_{ - \infty }^\infty  {\phi \left( x \right)^q \left\{ { - \frac{{\beta _q  ^{(g)} \left( {x - \mu_q } \right)^2 }}{{\sigma_q ^2 }} + \exp _q \left( { - \frac{{\beta _q  ^{(g)} \left( {x - \mu_q } \right)^2 }}{{\sigma_q ^2 }}} \right)^{1 - q} \ln _q \frac{1}{{Z_q^{(g)} }}} \right\}dx}  \\ 
 &&=  - \frac{{\beta _q^{(g)} }}{{\sigma_q ^2 }}\int_{ - \infty }^\infty  {\left( {x - \mu_q } \right)^2 \phi \left( x \right)^q dx}  + \ln _q \frac{1}{{Z_q^{(g)} }}\int_{ - \infty }^\infty  {\phi \left( x \right)^q \left\{ {1 - \frac{{\beta _q ^{(g)}  \left( {1 - q} \right)\left( {x - \mu_q } \right)^2 }}{{\sigma_q ^2 }}} \right\}dx}  \\ 
 &&=  - \beta _q ^{(g)}  c_q  + c_q \ln _q \frac{1}{{Z_q^{(g)} }} - \beta _q ^{(g)}  c_q \left( {1 - q} \right)\ln _q \frac{1}{{Z_q^{(g)} }} \\ 
 &&= - \beta _q  ^{(g)} c_q Z_q^{{(g)} ^{q - 1}}  + c_q \ln _q \frac{1}{{Z_q^{(g)} }}. 
 \end{eqnarray*}
Thus we have 
\[
H_q \left( {\phi \left( x \right)} \right) \equiv  - \int_{ - \infty }^\infty  {\phi \left( x \right)^q \ln _q \phi \left( x \right)dx} 
 \le  - \int_{ - \infty }^\infty  {\phi \left( x \right)^q \ln _q \psi \left( x \right)dx} \, = c_q \beta _q ^{(g)}  Z_q^{{(g)} ^{q - 1}}  - c_q \ln _q \frac{1}{{Z_q^{(g)} }}
\]
by the nonnegativity of the Tsallis relative entropy.
From the equality condition of the Tsallis relative entropy,
we see that the maximum attains if and only if
\[
\phi \left( x \right) = \psi \left( x \right) = \frac{1}{{Z_q^{(g)} }}\exp _q \left( { - \frac{{\beta _q ^{(g)}  \left( {x - \mu_q } \right)^2 }}{{\sigma_q ^2 }}} \right).
\]

\hfill \qed
\begin{Cor}\label{cor2}
If $\phi \in C_1^{(g)}$, then $H_1(\phi(x)) \leq \log \sqrt{2\pi e}\sigma$ with equality  if and only if  $\phi(x) = \frac{1}{\sqrt{2\pi}\sigma} \exp\left\{ -\frac{(x - \mu)^2}{2\sigma^2} \right\}$.
\end{Cor}
{\it Proof}:
Take the limit $q \to 1$ in Theorem \ref{the2}.
\hfill \qed

\section{Minimization of $q$-Fisher information}  \label{sec3}
The theorem in the previous section and the fact that the Gaussian distribution minimizes 
the Fisher information
lead us to study the Tsallis distribution 
($q$-Gaussian distribution) minimizes a $q$-Fisher information as a one-parameter extension.
We prepare some definitions for this purpose.
In what follows, we abbreviate $\beta_q$ and $Z_q$ instead of $\beta_q^{(g)}$ and $Z_q^{(g)}$, respectively.
\begin{Def}
For the random variable $X$ with the probability density function $f(x)$, we define the $q$-score function $s_q(x)$ and $q$-Fisher information $J_q(X)$ by
\begin{eqnarray}
&& s_q(x) \equiv \frac{d \ln_q f(x)}{dx}, \label{q_score}\\
&& J_q(X) \equiv E_q\left[s_q(x)^2\right],
\end{eqnarray}
where a normalized $q$-expectation value $E_q$ is defined by $E_q[g(X)] \equiv \frac{\int g(x)f(x)^qdx}{\int f(x)^qdx}$ for random variables $g(X)$ for any continuous function $g(x)$
and the probanility density function $f(x)$.
\end{Def}

Note that our definition of a $q$-Fisher information is different from those 
in several literature \cite{PPM,PPP1,Abe2,PP,PPP2,PPP3,Has}.


\begin{Ex}
For the random variable $G$ obeying to $q$-Gaussian distribution 
$$\phi_q^{(g)}(x)\equiv \frac{1}{Z_q}\exp_q\left\{-\frac{\beta_q\left(x-\mu_q\right)^2}{\sigma_q^2}\right\},$$
where $\beta_q\equiv \frac{1}{3-q}$ and $q$-partition function 
$Z_q\equiv \int_{-\infty}^{\infty} \exp_q\left\{ -\frac{\beta_q\left(x-\mu_q\right)^2}{\sigma_q^2} \right\}dx$,
the $q$-score function is calculated as 
$$
s_q(x) = -\frac{2\beta_qZ_q^{q-1}}{\sigma_q^2}\left(x-\mu_q\right).
$$
Thus we can calculate the $q$-Fisher information as
\begin{equation}   \label{qFisher1}
J_q(G)=\frac{4\beta_q^2Z_q^{2q-2}}{\sigma_q^2}.
\end{equation}
Note that we have
\begin{equation}   \label{Fisher_limit1}
\lim_{q\to 1}J_q(G) = \frac{1}{\sigma_1^2}.
\end{equation}
\end{Ex}
\begin{The}  \label{qFisher_theorem1}
Given the random variable $X$ with the probability density function $p(x)$, 
the $q$-expectation value $\mu_q \equiv E_q\left[X\right]$ and 
the $q$-variance $\sigma_q^2 \equiv E_q\left[\left(X-\mu_q\right)^2\right]$, 
we have a $q$-Cram\'er-Rao inequality:
\begin{equation}\label{qFisher2}
J_q(X) \geq \frac{1}{\sigma_q^2}\left(\frac{2}{\int p(x)^qdx}-1\right)\,\,\, for\,\,\,q\in [0,1) \cup (1,3).
\end{equation}
Immediately we have 
\begin{equation}\label{qFisher_Immed}
J_q(X)\geq \frac{1}{\sigma_q^2}\quad for \quad q \in (1,3).
\end{equation}
\end{The}
{\it Proof}:
Here we assume that $\lim_{x\to \pm \infty} f(x)p(x) = 0$ for any $q \geq 0$, any probability density function $p(x)$ and
 any smooth function $f$ which is suitably well-behaved at $\pm \infty$. 
Then we have 
\begin{eqnarray*}
E_q\left[  \left(X-\mu_q\right) s_q(x)  \right] &=& \frac{\int \left(x-\mu_q\right) p(x)^qs_q(x)dx}{\int p(x)^q dx}\\
&=&  \frac{\int \left(x-\mu_q\right) p'(x)dx}{\int p(x)^q dx}\\
&=& \frac{-1}{\int p(x)^qdx}.
\end{eqnarray*}
Thus we have
\begin{eqnarray*}
0 &\leq& E_q\left[\left\{ s_q(x)+ \frac{\left(X-\mu_q\right)}{\sigma_q^2} \right\}^2  \right]\\
&=& J_q(X)+\frac{2}{\sigma_q^2}E_q\left[ \left(X-\mu_q\right) s_q(x)   \right] +\frac{E_q\left[ \left(X-\mu_q\right)^2 \right]}{\sigma_q^4}\\
&=& J_q(X) -\frac{2}{\sigma_q^2\int p(x)^qdx} + \frac{1}{\sigma_q^2},
\end{eqnarray*}
which implies a $q$-Cram\'er-Rao lower bound given in (\ref{qFisher2}).
\hfill \qed

\begin{Prop}\label{qFisher_prop}
The equality in the $q$-Cram\'er-Rao inequality (\ref{qFisher2}) holds if
the probability density function $p(x)$ is the $q$-Gaussian density function $\phi_q^{(g)}(x)$ with
the $q$-variance 
\begin{equation}    \label{equality1}
 \sigma _q  = 
 \frac{{2^{\frac{q}{{1 - q}}} \left( {3 - q} \right)^{\frac{{q + 1}}{{2\left( {q - 1} \right)}}} \left( {1 - q} \right)^{\frac{1}{2}} }}{{B\left( {\frac{1}{2},\frac{1}{{1 - q}}} \right)}},\,\,\,\,\,\left( {0 \leq q < 1} \right) 
\end{equation}
or
\begin{equation}\label{equality2}
 \sigma _q  = 
 \frac{{2^{\frac{1}{{1 - q}}} \left( {3 - q} \right)^{\frac{{3 - q}}{{2\left( {q - 1} \right)}}} \left( {q - 1} \right)^{\frac{1}{2}} }}{{B\left( {\frac{1}{{q - 1}} - \frac{1}{2},\frac{1}{2}} \right)}},\,\,\,\,\,\left( {1 < q < 3} \right). 
\end{equation}
\end{Prop}
{\it Proof}:
We show that the following inequality holds for $0 \leq q < 3$ and $q \neq 1$ 
\begin{equation}   \label{qFisher3}
J_q(G) \geq \frac{1}{\sigma_q^2}\left(\frac{2}{\int \phi_q^{(g)}(x)^qdx}-1\right),
\end{equation}
with equality if the $q$-variance is given by (\ref{equality1}) or (\ref{equality2}).
\begin{itemize}
\item[(i)] For the case of $0 \leq q <1$,
 we firstly calculate 
\begin{eqnarray*}
&&   Z_q  \equiv \int_{ - \infty }^\infty  {\exp _q \left\{ { - \frac{{\beta _q \left( {x - \mu _q } \right)^2 }}{{\sigma _q^2 }}} \right\}dx} 
     = \int_{ - \infty }^\infty  {\exp _q \left\{ { - \frac{{\beta _q y^2 }}{{\sigma _q^2 }}} \right\}dy}  \\ 
&&   = 2\sigma _q \int_0^{\sqrt {\frac{{3 - q}}{{1 - q}}} } {\left( {1 - \frac{{1 - q}}{{3 - q}}z^2 } \right)} ^{\frac{1}{{1 - q}}} dz 
     = 2\sigma _q \sqrt {\frac{{3 - q}}{{1 - q}}} \int_0^1 {\left( {1 - t^2 } \right)^{\frac{1}{{1 - q}}} dt}\\  
&&   = \sigma _q \left( {\frac{{3 - q}}{{1 - q}}} \right)^{\frac{1}{2}} B\left( {\frac{1}{2},\frac{1}{{1 - q}} + 1} \right) 
 \end{eqnarray*}
and
\begin{eqnarray*}
&&\int_{ - \infty }^\infty  {\phi_q^{(g)} \left( x \right)^qdx}  = \frac{1}{{Z_q^q }}\int_{ - \infty }^\infty  {\exp _q \left\{ { - \frac{{\beta _q \left( {x - \mu _q } \right)^2 }}{{\sigma _q^2 }}} \right\}^q dx}  
  = \frac{{2\sigma _q }}{{Z_q^q }}\int_0^{\sqrt {\frac{{3 - q}}{{1 - q}}} } {\left( {1 - \frac{{1 - q}}{{3 - q}}z^2 } \right)} ^{\frac{q}{{1 - q}}} dz \\ 
&&  = \frac{{2\sigma _q }}{{Z_q^q }}\left( {\frac{{3 - q}}{{1 - q}}} \right)^{\frac{1}{2}} \int_0^1 {\left( {1 - t^2 } \right)^{\frac{q}{{1 - q}}} dt}  
  = \frac{{\sigma _q }}{{Z_q^q }}\left( {\frac{{3 - q}}{{1 - q}}} \right)^{\frac{1}{2}} B\left( {\frac{1}{2},\frac{1}{{1 - q}}} \right). 
\end{eqnarray*}
Then the L.H.S. and the R.H.S. of (\ref{qFisher3}) are calculated as
\[
2^{2q} \sigma _q ^{2q - 2} \left( {3 - q} \right)^{ - \left( {q + 1} \right)} \left( {1 - q} \right)^{1 - q} B\left( {\frac{1}{2},\frac{1}{{1 - q}}} \right)^{2q - 2} 
\]
and
\[
2^{q + 1} \sigma _q ^{q - 1} \left( {3 - q} \right)^{ - \frac{{q + 1}}{2}} \left( {1 - q} \right)^{\frac{{1 - q}}{2}} B\left( {\frac{1}{2},\frac{1}{{1 - q}}} \right)^{q - 1}  - 1,
\]
respectively.
Then we have the inequality
\begin{eqnarray*}
&& {\rm{L}}{\rm{.H}}{\rm{.S}}{\rm{.}} - {\rm{R}}{\rm{.H}}{\rm{.S}}{\rm{.}} \\ 
&&  = 2^{2q} \sigma _q ^{2q - 2} \left( {3 - q} \right)^{ - \left( {q + 1} \right)} \left( {1 - q} \right)^{1 - q} B\left( {\frac{1}{2},\frac{1}{{1 - q}}} \right)^{2q - 2}\\
&&  - 2^{q + 1} \sigma _q ^{q - 1} \left( {3 - q} \right)^{ - \frac{{q + 1}}{2}} \left( {1 - q} \right)^{\frac{{1 - q}}{2}} B\left( {\frac{1}{2},\frac{1}{{1 - q}}} \right)^{q - 1}  + 1 \\ 
&&  = \left\{ {2^q \sigma _q ^{q - 1} \left( {3 - q} \right)^{ - \frac{{q + 1}}{2}} \left( {1 - q} \right)^{\frac{{1 - q}}{2}} B\left( {\frac{1}{2},\frac{1}{{1 - q}}} \right)^{q - 1}  - 1} \right\}^2  \ge 0,  
\end{eqnarray*}
with equality if Eq.(\ref{equality1}) holds.
\item[(ii)] For the case of $1 <q <3$,
 we similarly calculate 
\[
Z_q  = \sigma _q \left( {\frac{{3 - q}}{{q - 1}}} \right)^{\frac{1}{2}} B\left( {\frac{1}{{q - 1}} - \frac{1}{2},\frac{1}{2}} \right)
\]
and
\[
\int_{ - \infty }^\infty  {\phi_q^{(g)} \left( x \right)^qdx}  = \frac{{\sigma _q }}{{Z_q ^q }}\left( {\frac{{3 - q}}{{q - 1}}} \right)^{\frac{1}{2}} B\left( {\frac{q}{{q - 1}} - \frac{1}{2},\frac{1}{2}} \right).
\]
Then the L.H.S. and the R.H.S. of (\ref{qFisher3}) are calculated as
\[
4\sigma _q ^{2q - 2} \left( {3 - q} \right)^{q - 3} \left( {q - 1} \right)^{1 - q} B\left( {\frac{1}{{q - 1}} - \frac{1}{2},\frac{1}{2}} \right)^{2q - 2} 
\]
and
\[
4\sigma _q ^{q - 1} \left( {3 - q} \right)^{\frac{{q - 3}}{2}} \left( {q - 1} \right)^{\frac{{1 - q}}{2}} B\left( {\frac{1}{{q - 1}} - \frac{1}{2},\frac{1}{2}} \right)^{q - 1}  - 1,
\]
respectively.
Then we have the inequality
\begin{eqnarray*}
 &&{\rm{L}}{\rm{.H}}{\rm{.S}}{\rm{.}} - {\rm{R}}{\rm{.H}}{\rm{.S}}{\rm{.}} \\ 
 &&= 4\sigma _q ^{2q - 2} \left( {3 - q} \right)^{q - 3} \left( {q - 1} \right)^{1 - q} B\left( {\frac{1}{{q - 1}} - \frac{1}{2},\frac{1}{2}} \right)^{2q - 2}  \\ 
 && - 4\sigma _q ^{q - 1} \left( {3 - q} \right)^{\frac{{q - 3}}{2}} \left( {q - 1} \right)^{\frac{{1 - q}}{2}} B\left( {\frac{1}{{q - 1}} - \frac{1}{2},\frac{1}{2}} \right)^{q - 1}  + 1 \\ 
 && = \left\{ {2\sigma _q ^{q - 1} \left( {3 - q} \right)^{\frac{{q - 3}}{2}} \left( {q - 1} \right)^{\frac{{1 - q}}{2}} B\left( {\frac{1}{{q - 1}} - \frac{1}{2},\frac{1}{2}} \right)^{q - 1}  - 1} \right\}^2  \ge 0, 
\end{eqnarray*}
with equality if Eq.(\ref{equality2}) holds.
\end{itemize}
\hfill \qed

Note that we have $J_1(X) \geq \frac{1}{\sigma_1^2}$ in the limit $q\to 1$.
Proposition \ref{qFisher_prop} also shows that $q$-Gaussian with $q$-variance such that Eq.(\ref{equality1}) or Eq.(\ref{equality2})
minimizes the $q$-Fisher information. In addition, we note on the limit $q\to 1$ for 
the $q$-variances $\sigma_q$ given in Eq.(\ref{equality1}) and Eq.(\ref{equality2}).
The following results were checked by the computer software:
$$
\lim_{q\to 1-0} \sigma_q 
=  \lim_{q\to 1-0} \frac{{2^{\frac{q}{{1 - q}}} \left( {3 - q} \right)^{\frac{{q + 1}}{{2\left( {q - 1} \right)}}} \left( {1 - q} \right)^{\frac{1}{2}} }}{{B\left( {\frac{1}{2},\frac{1}{{1 - q}}} \right)}} 
= \lim_{r\to +0} \frac{2^{\frac{1-r}{r}} (2+r)^{\frac{r-2}{2r}}r^{\frac{1}{2}}}{B\left(\frac{1}{2},\frac{1}{r}\right)}
= \frac{1}{\sqrt{2 e \pi}}
$$
and 
$$
\lim_{q\to 1+0} \sigma_q 
= \lim_{q \to 1+0}  \frac{{2^{\frac{1}{{1 - q}}} \left( {3 - q} \right)^{\frac{{3 - q}}{{2\left( {q - 1} \right)}}} \left( {q - 1} \right)^{\frac{1}{2}} }}{{B\left( {\frac{1}{{q - 1}} - \frac{1}{2},\frac{1}{2}} \right)}}
= \frac{1}{\sqrt{2 e \pi}}.
$$


\begin{Rem}
In our previous paper \cite{Furu}, we gave the rough meaning of the parameter 
$q$ from the information-theoretical viewpoint. 
In \cite{Furu}, we showed that the Tsallis entropies for $q \geq 1$ had the subadditivity 
and therefore we had several information-theoretical properties 
in the case of $q \geq 1$. However,  the Tsallis entropies for $q < 1$ did not have such properties. 
As similar as the case of the Tsallis entropies, in the present paper 
we have found that $q$-Fisher information have the quite same situation 
such that we have $J_q(X) \geq \frac{1}{\sigma_q^2}$ for $q \geq 1$,
however for the case of $q <1$, we do not have any relation between $J_q(X)$ and $\frac{1}{\sigma_q^2}$ other than the inequality (\ref{qFisher2}).
Therefore these results give us the difference of the $q$-Fisher information $J_q(X)$ for $q \in [0,1)$ 
and $J_q(X)$ for $q \in (1,3)$,
as similar as the Tsallis entropies did in \cite{Furu}.
Summarizing these results, we may conclude that the Tsallis entropies 
and $q$-Fisher information make a sense for the case of $q \geq 1$ in our setting.
\end{Rem}


\section{Concluding remarks} \label{sec4}
Throughout the present paper, we adopted the normalized $q$-expectation value 
as a one-parameter generalization of the standard expectation value.
In this section, we consider on our results obtained in Section \ref{sec2} and \ref{sec3}, 
for different expectation values.
The normalized $q$-expectation value 
$E_q[X]\equiv \frac{\int xf(x)^qdx}{\int f(x)^qdx}$
adopted in the present paper has mathematical desirable properties so that it was used in many literatures on Tsallis statistics,
and 
is rewritten by the standard expectation value as 
$E_1[X]\equiv \int xh(x)dx$ where $h(x)\equiv \frac{f(x)^q}{\int f(x)^qdx}$ is often called the escort density function.

If we adopt the constraints $C_1^{(c)}$ or $C_1^{(g)}$ 
due to the standard expectation value $E_1[X]\equiv \int xf(x)dx$,
Theorem \ref{the1} and Theorem \ref{the2} can not be derived by the use of nonnegativity of the Tsallis relative entropy,
 as easily seen from the processes of their proofs. However for the standard expectation value,
 we have Corollary \ref{cor1} and Corollary \ref{cor2}.  
That is, it was reconfirmed that the standard expectation value $E_1$ corresponds to Shannon entropy and Kullback-Leibler information.

As a one-parameter generalization of the standard expectation value $E_1$, 
the following $q$-expectation value may be considered:
$$\widetilde{E_q}[X] \equiv \int_{-\infty}^{\infty} xf(x)^qdx. $$
For this $q$-expectation value, we also have the following results.
Define the constraints:
\begin{eqnarray*}
&&  \widetilde{C_q^{(c)}}\equiv \left\{f\in D : \int_{-\infty}^{\infty} xf(x)^qdx= \mu_q \right\}, \\
&&  \widetilde{C_q^{(g)}}\equiv \left\{f\in \widetilde{C_q^{(c)}} : \int_{-\infty}^{\infty} (x-\mu_q)^2f(x)^qdx= \sigma_q^2 \right\}. 
\end{eqnarray*} 
Then we have the following results by the similar way to Theorem \ref{the1} and Theorem \ref{the2}.
\begin{The}  \label{the4_1}
\begin{itemize}
\item[(1)] If $\phi \in \widetilde{C_q^{(c)}}$, then
$$H_q(\phi(x)) \leq -c_q \ln_q \frac{1}{Z_q^{(c)}},$$
with equality if and only if
$$\phi(x) = \frac{1}{Z_q^{(c)}}\exp_q\left\{-\beta_q^{(c)}\left(x-\mu_q\right)\right\},$$
where $\beta_q^{(c)}$ and $Z_q^{(c)}$ are same constant numbers in Theorem \ref{the1}.
\item[(2)]For $q\in\mathbb{R}$ such that $0\leq q<3$ and $q\neq 1$ if $\phi \in \widetilde{C_q^{(g)}}$, 
then $$H_{q}(\phi(x)) \leq  -c_q \ln_q \frac{1}{Z_q^{(g)}} +\beta_q^{(g)} Z_q^{{(g)}^{q-1}}     ,$$ with equality
 if and only if  $$\phi(x) = \frac{1}{Z_q^{(g)}}     \exp_q\left( - \frac{\beta_q  ^{(g)} {(x - \mu_q)^2}}{\sigma_q^2}  \right)    ,$$ 
where $\beta_q^{(g)}$ and $Z_q^{(g)}$ are same constant numbers in Theorem \ref{the2}.
\end{itemize}
\end{The}
Moreover, we may define a $q$-Fisher information by the $q$-expectation value $\widetilde{E_q}[X]$ 
as $\widetilde{J_q}(X)\equiv\widetilde{E_q}[s_q(x)^2]$,
where $s_q(x)$ is a same score function in Eq.(\ref{q_score}).
Then we have the following result by the similar way to Theorem \ref{qFisher_theorem1}.
\begin{The}  \label{the4_2}
Given the random variable $X$ with the probability density function $p(x)$, 
the $q$-expectation value $\mu_q \equiv \widetilde{E_q}\left[X\right]$ and 
the $q$-variance $\sigma_q^2 \equiv \widetilde{E_q}\left[\left(X-\mu_q\right)^2\right]$, 
we have a $q$-Cram\'er-Rao inequality:
\begin{equation}\label{non_normalized_CR}
\widetilde{J_q}(X) \geq \frac{1}{\sigma_q^2} \quad for\,\,\,q\in [0,1) \cup (1,3).
\end{equation}
In addition, the equality holds if $p(x)=\phi_q^{(g)}(x)$ and $\sigma_q$ are given by Eq.(\ref{equality1}) or Eq.(\ref{equality2}).
\end{The}
Thus we can see that Theorem \ref{the4_1} and Theorem \ref{the4_2} are almost similar to the results obtained 
in Thorem \ref{the1}, Theorem \ref{the2} and Theorem \ref{qFisher_theorem1}, except for the normalization factor.
We also find that Theorem \ref{the4_2} has a slightly modified form, if it is compared with Theorem \ref{qFisher_theorem1},
because a one-parameter generalized Cram\'er-Rao inequality (\ref{non_normalized_CR}) 
holds for any $q \in \mathbb{R}$ such that $0 \leq q < 3$ and $q\neq 1$,
while the inequality (\ref{qFisher_Immed}) holds for $1< q < 3$.

We close this section giving a comment on a possible application of our $q$-Fisher informations.
The central limit theorem, which is one of important theorems in probability theory, 
states the distribution function of the standardized sum of an independent sequence
of random variables convergences to Gaussian distribution under a certain assumpution.
The classical central limit theorem is usually proved by the characteristic function.
However it is known that the Fisher information can be applied to prove the classical central limit theorem \cite{Bro,Shi,Joh}.
In addition, quite recently, the $q$-central limit theorem for $q \geq 1$ was proved in \cite{UTS} 
by introducing new notions such as $q$-independence, $q$-convergence,
$q$-Fourier transformation and $q$-characteristic function.
Therefore we may expect that a new proof of $q$-central limit theorem may be given by applying $q$-Fisher information in the future.

\section*{Acknowledgement}
This work was supported by the Japanese Ministry of Education, Science, Sports and Culture, Grant-in-Aid for 
Encouragement of Young Scientists (B), 20740067 and Grant-in-Aid for Scientific Research (B), 18300003.

\newpage

\section*{Appendix A}
On Theorem \ref{the2}, we here show $q$-Gaussian distribution:
\begin{equation}
\phi(x) = Z_q^{-1}  \exp_q\left\{ -\beta_q (x - \mu_q)^2/{\sigma_q^2}  \right\},
\end{equation}
 where $   Z_q \equiv \int_{-\infty}^{\infty}  \exp_q\left\{ -\beta_q (x - \mu_q)^2/{\sigma_q^2}  \right\}dx$ with $\beta_q = 1/(3-q)$ satisfies the constraints:
\begin{equation}
\frac{1}{c_q}\int_{-\infty}^{\infty}x \phi(x)^qdx = \mu_q    
\end{equation}
and
\begin{equation}
\frac{1}{c_q}  \int_{-\infty}^{\infty}(x - \mu_q)^2\phi(x)^qdx = \sigma_q^2,     
\end{equation}
where $c_q \equiv \int_{-\infty}^{\infty} \phi(x)^qdx$. 

{\it Proof}:
It is sufficient to prove the case of $\mu_q = 0$ and $\sigma_q = 1$.
Since the function 
$x\exp _q \left( { - \frac{{x^2 }}{{\left( {3 - q} \right)}}} \right)^q $
 is the odd function, we see that $\phi(x)$ satisfies the first constraint. 
To show the second constraint is equivalent to show
\begin{equation} \label{appen1}
\int_{-\infty}^{\infty}x^2 \exp_q\left( -\frac{x^2}{(3-q)} \right)^q dx =   \int_{-\infty}^{\infty} \exp_q \left( -\frac{x^2}{(3-q)} \right)^q dx.      
\end{equation}

\begin{itemize}
\item[(1)] $0\leq q <1$: 
Since
\[
\exp _q \left( { - \frac{{x^2 }}{{\left( {3 - q} \right)}}} \right) = \left\{ \begin{array}{l}
 \left( {1 - \frac{{\left( {1 - q} \right)x^2 }}{{\left( {3 - q} \right)}}} \right)^{\frac{1}{{1 - q}}} \,\,\,\,\,\,\,\,\,\,\,\,\,\,\,\,\,\,\,\,\,\,\,\,\,\,\,\,if\,\,\,\, - \sqrt {\frac{{3 - q}}{{1 - q}}}  < x < \sqrt {\frac{{3 - q}}{{1 - q}}}  \\ 
 \,\,\,\,\,\,\,\,\,\,\,\,\,\,\,\,\,\,\,\,\,\,\,\,0\,\,\,\,\,\,\,\,\,\,\,\,\,\,\,\,\,\,\,\,\,\,\,\,\,\,\,\,\,\,\,\,\,\,\,\,\,\,\,\,\,\,\,\,\,\,\,otherwise \\ 
 \end{array} \right.
\]
The L.H.S. of Eq.(\ref{appen1}) is calculated by
\begin{eqnarray*}
 \int_{ - \sqrt {\frac{{3 - q}}{{1 - q}}} }^{\sqrt {\frac{{3 - q}}{{1 - q}}} } {x^2 \exp _q \left( { - \frac{{x^2 }}{{\left( {3 - q} \right)}}} \right)^q dx}  &=& 
\left( {\frac{{3 - q}}{{1 - q}}} \right)^{3/2} \int_{ - 1}^1 {y^2 \left( {1 - y^2 } \right)^{\frac{q}{{1 - q}}} dy}  \\ 
  &=& 2\left( {\frac{{3 - q}}{{1 - q}}} \right)^{3/2} \int_0^1 {y^2 \left( {1 - y^2 } \right)^{\frac{q}{{1 - q}}} dy} \\
 &=& \left( {\frac{{3 - q}}{{1 - q}}} \right)^{3/2} B\left( {\frac{3}{2},\frac{1}{{1 - q}}} \right). 
 \end{eqnarray*}
Also the R.H.S. of Eq.(\ref{appen1}) is calculated by
\begin{eqnarray*}
 \int_{ - \sqrt {\frac{{3 - q}}{{1 - q}}} }^{\sqrt {\frac{{3 - q}}{{1 - q}}} } {\exp _q \left( { - \frac{{x^2 }}{{\left( {3 - q} \right)}}} \right)^q dx}  &=&
 \left( {\frac{{3 - q}}{{1 - q}}} \right)^{1/2} \int_{ - 1}^1 {\left( {1 - y^2 } \right)^{\frac{q}{{1 - q}}} dy}  \\ 
  &=& 2\left( {\frac{{3 - q}}{{1 - q}}} \right)^{1/2} \int_0^1 {\left( {1 - y^2 } \right)^{\frac{q}{{1 - q}}} dy} \\
 &=&  \left( {\frac{{3 - q}}{{1 - q}}} \right)^{1/2} B\left( {\frac{1}{2},\frac{1}{{1 - q}}} \right). 
 \end{eqnarray*}
In the process of the above calculations, the following formula:
\[
\int_0^1 {x^{p-1}  \left( {1 - x^a  } \right)^{r-1}  dx}  
= \frac{1}{a}B\left( {\frac{{p}}{a},r} \right),\,\,\,\,\,\,\left( {a>0 ,p>0, r>0} \right)
\]
was used. By the properties of the beta function and gamma function, 
$\left( {\frac{{3 - q}}{{1 - q}}} \right)^{3/2} B\left( {\frac{3}{2},\frac{1}{{1 - q}}} \right)$ 
 coincides with 
$\left( {\frac{{3 - q}}{{1 - q}}} \right)^{1/2} B\left( {\frac{1}{2},\frac{1}{{1 - q}}} \right)$.
\item[(2)] $1<q\leq 3$:
The L.H.S. of Eq.(\ref{appen1}) is calculated by
\begin{eqnarray*}
 \int_{ - \infty }^\infty  {x^2 \exp _q \left( { - \frac{{x^2 }}{{\left( {3 - q} \right)}}} \right)^q dx}  &=& 
2\left( {\frac{{3 - q}}{{q - 1}}} \right)^{3/2} \int_0^\infty  {y^2 \left( {1 + y^2 } \right)^{\frac{q}{{1 - q}}} dy}  \\ 
  &=& \left( {\frac{{3 - q}}{{q - 1}}} \right)^{3/2} B\left( {\frac{q}{{q - 1}} - \frac{3}{2},\frac{3}{2}} \right). 
\end{eqnarray*}
The R.H.S. of Eq.(\ref{appen1}) is calculated by
\begin{eqnarray*}
 \int_{ - \infty }^\infty  {\exp _q \left( { - \frac{{x^2 }}{{\left( {3 - q} \right)}}} \right)^q dx}  &=& 
2\left( {\frac{{3 - q}}{{q - 1}}} \right)^{1/2} \int_0^\infty  {\left( {1 + y^2 } \right)^{\frac{q}{{1 - q}}} dy}  \\ 
  &=& \left( {\frac{{3 - q}}{{q - 1}}} \right)^{1/2} B\left( {\frac{q}{{q - 1}} - \frac{1}{2},\frac{1}{2}} \right). 
\end{eqnarray*}
In the process of the above calculations, the following formula:
\[
\int_0^\infty  {\frac{{dx}}{{x^p  \left( {1 + x^a  } \right)^r  }}}  
= \frac{1}{a}B\left( {r  + \frac{{p-1 }}{a},\frac{{1 - p}}{a}} \right),
\quad \left( {a>0,p  < 1,r  > 0,ar> 1 - p} \right),
\]
which is derived from $B(p,r)=B(r,p)$ and 
$$B(p,r)=\int_0^{\infty}\frac{x^{p-1}}{\left(1+x\right)^{p+r}}dx,\quad \left(p>0,r>0\right),$$
was used.
By the properties of the beta function and gamma function, 
$\left( {\frac{{3 - q}}{{q - 1}}} \right)^{3/2} B\left( {\frac{q}{{q - 1}} - \frac{3}{2},\frac{3}{2}} \right)$
 coincides with 
$\left( {\frac{{3 - q}}{{q - 1}}} \right)^{1/2} B\left( {\frac{q}{{q - 1}} - \frac{1}{2},\frac{1}{2}} \right).$
\end{itemize}

\section*{Appendix B}
{\it Proof of Theorem \ref{the4_2}}:
The inequality (\ref{non_normalized_CR}) follows by the similar way to the proof 
of Theorem \ref{qFisher_theorem1} so that we prove the equality condition.
We have $\widetilde{J_q}(G)=\frac{4\beta_q^2Z_q^{2q-2}}{\sigma_q}$ for the $q$-Gaussian density function $\phi_q^{(g)}(x)$. 
In addition, we see that the equation $4\beta_q^2Z_q^{2q-2}=1$ is equivalent to Eq.(\ref{equality1}) or Eq.(\ref{equality2}),
thanks to the formula $B(p,r+1)=\frac{r}{p+r}B(p,r)$. 
\hfill \qed

\end{document}